# Nesting between hole and electron pockets in Ba(Fe$_{1-x}$Co$_x$)$_2$As$_2$ ($x$=0-0.3) observed with angle-resolved photoemission

V. Brouet[1], M. Marsi[1], B. Mansart[1], A. Nicolaou[1,2], A. Taleb-Ibrahimi[2], P. Le Fèvre[2], F. Bertran[2],
F. Rullier-Albenque[3], A. Forget[3] and D. Colson[3]

[1] *Lab. Physique des Solides, Université Paris-Sud, UMR8502, Bât 510, 91405 Orsay, France*
[2] *Synchrotron SOLEIL, L'Orme des Merisiers, Saint-Aubin-BP48, 91192 Gif-sur-Yvette, France*
[3] *Service de Physique de l'Etat Condensé, Orme des Merisiers, CEA Saclay (CNRS URA 2464), 91191 Gif sur Yvette cedex, France*

We present a comprehensive angle-resolved photoemission study of the three-dimensional electronic structure of Ba(Fe$_{1-x}$Co$_x$)$_2$As$_2$. The wide range of dopings covered by this study, $x$=0 to $x$=0.3, allows to extract systematic features of the electronic structure. We show that there are three different hole pockets around the Γ point, the two inner ones being nearly degenerate and rather two dimensional, the outer one presenting a strong three dimensional character. The structure of the electron pockets is clarified by studying high doping contents, where they are enlarged. They are found to be essentially circular and two dimensional. From the size of the pockets, we deduce the number of holes and electrons present at the various dopings. We find that the net number of carriers is in good agreement with the bulk stoichiometry, but that the number of each species (holes and electrons) is smaller than predicted by theory. Finally, we discuss the quality of nesting in the different regions of the phase diagram. The presence of the third hole pocket significantly weakens the nesting at $x$=0, so that it may not be a crucial ingredient in the formation of the Spin Density Wave. On the other hand, superconductivity seems to be favored by the coexistence of two-dimensional hole and electron pockets of similar sizes.

## I. INTRODUCTION

The newly discovered iron pnictides superconductors[1,2] are characterized by small hole and electron pockets, containing few carriers[3,4]. This semi-metallic band structure is rather unusual and contrasts with that of many correlated systems, where exotic properties are mostly found near half-filling of the bands. Many of the different properties observed in the iron pnictides family are thought to depend crucially on the interactions between these pockets. For the magnetic phases ($x$=0 and low dopings, see phase diagram on Fig. 7), the good nesting between hole and electron pockets should stabilize the spin-density wave ground state. For the superconducting phases, a non-conventional mechanism based on the exchange of spin fluctuations, with an order parameter changing sign between the hole and electron pockets has been put forward[4]. Angle-resolved photoemission spectroscopy is the best suited tool to observe the shape and size of the different pockets and how they evolve with doping, thanks to its unique ability to map the electronic structure in the reciprocal space. This is an exciting challenge that has triggered a tremendous activity in the past year[5-21].

It was soon confirmed that the electronic structure is made out of hole pockets at the Γ point and electron pockets at the Brillouin Zone (BZ) corners, as predicted from band structure calculations. This was shown for the so-called 1111 family, in LaFeOP[8] and NdFeAsO$_{1-x}$F$_x$[10], and, much more extensively due to the higher sample quality, for the 122 family, AFe$_2$As$_2$ (A=Ba, Sr,Ca). In this family, the size of these pockets was found to change with hole[11] or electron[18] doping, confirming one can transfer rather simply electron and holes to the Fe bands. For the hole-doped family Ba$_{1-x}$K$_x$Fe$_2$As$_2$ two different hole pockets were clearly resolved. They exhibit different superconducting gaps, evidencing that these bands may have different roles in the electronic properties[5]. However, there are still missing pieces of the puzzle to establish a global view of the electronic structure of these systems. In particular, the number of hole pockets, their degeneracy, the shape of the electron pockets, the two dimensional (2D) or three dimensional (3D) character of the different bands, are not totally clarified. This task is actually not trivial, as the bands often overlaps and may be difficult to resolve. In this paper, we establish these facts as clearly as possible for the electron-doped family Ba(Fe$_{1-x}$Co$_x$)$_2$As$_2$. Our aim is to discuss the quality of nesting between the different pockets as a function of $x$ and compare this to the evolution of electronic properties in the phase diagram.

$Ba(Fe_{1-x}Co_x)_2As_2$ has not yet been as extensively studied as $Ba_{1-x}K_xFe_2As_2$. Sekiba et al. [18] reported Fermi Surface (FS) for $x=0.15$ and Terashima et al.[19] the opening of a superconducting gap for $x=0.075$. Vilmercati et al. showed that the hole pocket at $x=0.1$ exhibits strong photon energy dependence[20]. Such a photon energy dependence is suggestive of 3D effects, as an ARPES experiment at a fixed photon energy maps the electronic structure at one particular $k_Z$ value[22]. If there is significant dispersion as a function of $k_Z$ (i.e. perpendicularly to the FeAs slab), measurements at different photon energies map inequivalent sections of the electronic structure and it will be necessary to study this dependence to get a correct three dimensional map of the pockets. We find that it is indeed the case here and report detailed measurements for a wide range of photon energies (20-100eV) and dopings : $x=0$ ($T_{SDW}=139K$), $x=0.045$ ($T_{SDW}=63K$, $T_c=12K$), $x=0.08$ ($T_c=23K$), $x=0.15$ ($T_c=0$) and $x=0.3$ ($T_c=0$). We show that a common electronic structure with 3 hole pockets, 2 electron pockets, and a Fermi level adjusted to doping, describes well all cases. The structure of the hole pockets involves two nearly degenerate and 2D sheets and one larger sheet with stronger 3D character. While 3D effects were first thought negligible in K-doped $BaFe_2As_2$[14,6], there is increasing evidence that they are important at $x=0$ and for Co-doped compounds[20,15,12,21]. However, the fine structure of these 3D effects on the hole pockets is not clarified. Different models have been proposed[12,21] that we will discuss. On the other hand, we find that the two electron pockets are nearly degenerate and rather 2D. We characterize the holes and electrons Fermi velocities and effective mass, which are quite similar. Finally, we show that the structure we determine extrapolates well with that of the hole-doped side[7], which also clarifies the degeneracy and origin of these pockets.

With the topology of the FS in place, we extract the evolution of $k_F$ as a function of $x$. This gives a synthetic view (Fig. 7) of the correspondence between the pocket sizes and the different ground states. We also quantify the number of holes and electrons in each pocket (Fig. 8). We discuss these quantities in connection with the evolution of the electronic properties.

Before proceeding with this study, we recall important features of the electronic structure established by band calculations. The electronic structure is complicated by the fact that the five Fe 3d bands play an important role near the Fermi level and that there are 2 Fe per BZ (Fig. 1b), meaning actually 10 bands. To get a qualitative understanding of the structure, it is helpful to start with simplified models. As a first approximation, a 2 band model in a BZ containing 1 Fe (Fig. 1a) allows to describe the basic features[23]. The relevant orbitals are the Fe $3d_{XZ}$ and $3d_{YZ}$ orbitals, which are oriented along the diagonals of a Fe square and couple through As. They create a hole pocket and an electron pocket, both with mixed $d_{XZ}$ and $d_{YZ}$ character and of exactly the same size at $x=0$. These pockets then exhibit perfect nesting at $x=0$.

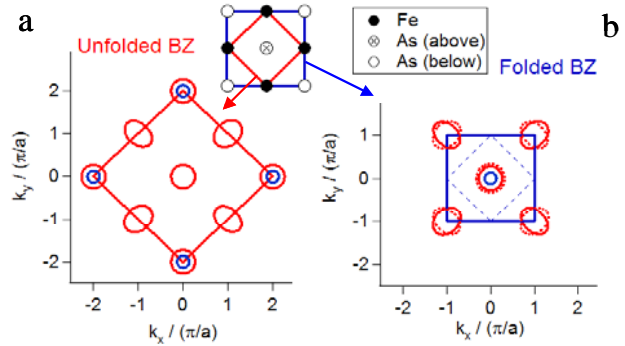

FIG. 1 : Sketch of the Brillouin Zones (BZ) corresponding to a Fe square (a) or the unit cell (b). Red circles (ovals) indicate 2D hole (electron) pockets, and blue circle the third hole pocket (see text). Dotted contours in (b) show the folded parts. The dashed blue square represents the AF BZ. Top inset : atoms in real space, a is the side of the unit cell (a=3.96Å).

When the 5 bands are included, there is significant rehybridization between all of them. The electron pockets for example acquire significant $d_{XY}$ character and a slightly more oval shape. More strikingly, bands of mainly $d_{XY}$ or $d_Z^2$ symmetry are found very near the Fermi level, and may even form an additional hole pocket (sketched in blue on Fig. 1). This disrupts the symmetry between the hole and electron pockets and therefore destroys the perfect nesting for the undoped system. Moreover, this pocket appears extremely sensitive to the structure of the FeAs slab, particularly the As height, and can be more 2D or 3D depending on its character[24]. It also seems to be different in the 1111 or 122 families, Nekrasov et al. predict it to be smaller than the other 2D hole pockets in LaFeAsO and larger in $BaFe_2As_2$[25]. Other studies predict it to be already filled in $BaFe_2As_2$[26]. It is then an important parameter of the electronic structure that is difficult to predict theoretically and should be determined experimentally.

Two different BZ can be considered, the "unfolded BZ" based on the Fe square (1 Fe/BZ) or the "folded BZ" based on the true structural unit cell (2Fe/BZ). The structural unit cell is twice as large as the Fe square to account for the inequivalent positions of the As above or below the Fe plane (see Fig. 1). In this paper, we quote all results in the folded BZ. The electronic structure in the folded BZ can essentially be obtained by folding all bands with respect to the new BZ boundaries (Fig. 1b). One ends up with 3 hole pockets at the Γ point and 2 electron pockets at the BZ corners. One can expect the two bands arising from the folding to be nearly degenerate and mostly 2D, while the third hole pocket may be more 3D.

Single crystals were grown using a FeAs and CoAs self-flux method[27]. All experiments were carried out at the CASSIOPEE beamline from the SOLEIL synchrotron, with a Scienta-R4000 analyser an angular resolution of 0.2° and an energy resolution better than 15meV. Samples were cleaved in ultra-high vacuum (better than $4.10^{-10}$ mbar), which exposes flat and shiny surfaces. The sample were found to evolve with time and fine structures could only be well resolved in the first 3 hours of measurement.

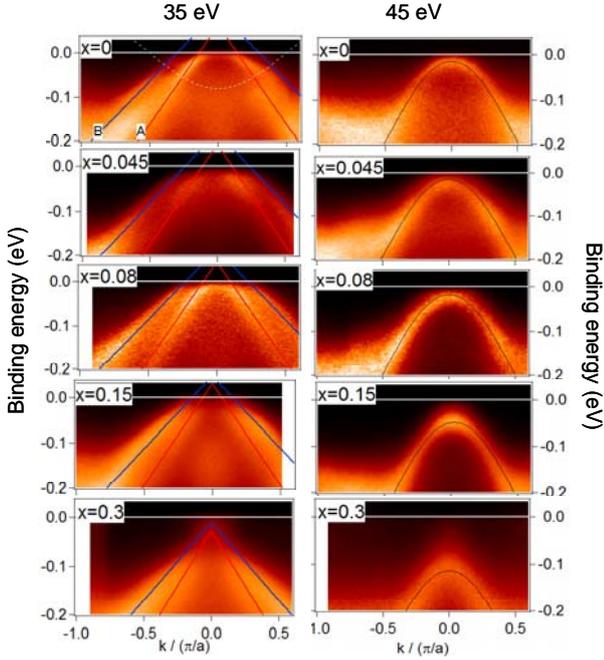

FIG. 2 : Near $E_F$ electronic structure around the $\Gamma$ point for the indicated dopings, measured at 35eV (left) or 45eV (right) photon energy and 30K. On the left part, lines sketch the dispersion of band A (red) and B (blue). On the right part, black line is a sketch for a cosine dispersion (see text).

## II. 3D STRUCTURE OF THE HOLE POCKETS

Fig. 2 presents the hole band structure for different $x$ values at 35eV (left) and 45eV (right). The data were taken at 30K, hence in the SDW for $x=0$ and $x=0.045$, but we do not consider here the changes associated to the SDW, which will be briefly discussed in section 5. Surprisingly, the band structure appears extremely different at the two photon energies, which is very unusual, and this has to be understood before anything can be said on the band structure. One would for example conclude that the hole pocket is already almost filled at $x=0$ looking at the 45eV data, and that it is not yet filled at $x=0.15$, looking at the 35eV data !

At 35eV, two bands are clearly distinguished, forming two hole pockets with distinct Fermi velocities and Fermi crossings. Their dispersion is sketched by red and blue lines, for bands called hereafter A and B, respectively. These two hole pockets have a nearly circular shape, as shown by the FS of Fig. 3a for $x=0.08$. When $x$ increases, the electron doping raise the position of the Fermi level in the band and the hole pockets shrink. The dispersions can be quite well fitted for all $x$ values, with linear dispersions keeping the same Fermi velocities, $V_F^A$=0.57eV.Å for the inner band A and $V_F^B$=0.41eV.Å for the outer band B. These fits are shown in Fig. 2 and used to extract $k_F$ values in Fig. 7.

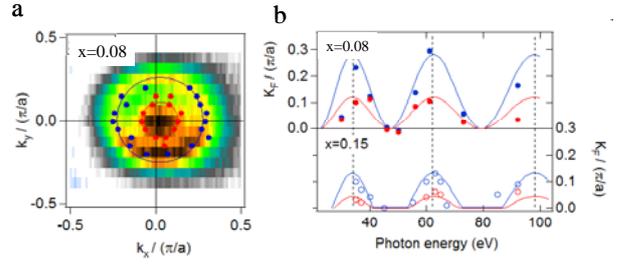

FIG. 3 : (a) Fermi Surface measured at 35eV and 30K, for $x=0.08$. Red and blue points indicate Fermi level crossings and the circles, the corresponding FS. (b) Fermi wave vector for the two bands as a function of photon energy for $x=0.08$ and $x=0.15$. Vertical dotted line show positions corresponding to $k_Z=1$. Solid lines are guides for the eye.

In sharp contrast, only one band is apparently observed at 45eV, with a more rounded shape. At $x=0$, it barely touches the Fermi level and sinks below it for higher $x$. Here again, the shape is roughly similar for all $x$. We describe it by a cosine function E=0.18cos(k*a), and shift it by 0.1eV between $x=0$ and $x=0.3$.

There are *a priori* two possible ways to explain the difference between 35 and 45eV. The first one is that the intensity of band B is strongly suppressed at 45eV by matrix element effects, so that it becomes undetectable and that band A dominates the spectrum. Note, however, that the dispersion at 45eV is already quite different from that of band A at 35eV (especially, the region at k=0 appears filled). The second one is to assume that the two bands have shifted with photon energy and have merged together at 45eV. This is possible if there is significant dispersion of the electronic structure perpendicularly to the surface, as different photon energies correspond to different $k_z$ values. These two possibilities would have very different implications for the understanding of the electronic structure. For example, the number of holes contained in these pockets would be quite different. They can be distinguished by observing these dispersions over a large photon energy window. If this behavior is related to perpendicular dispersion, it should display *oscillations* with a well defined periodicity related to $k_Z$, whereas no such periodicity is automatically expected for matrix-element effects. In Fig. 3b, the $k_F$ positions as a function of photon energy are shown for $x=0.08$ and $x=0.15$. They were extracted by linear fits, similar to those of Fig. 2. They indeed display well defined oscillations as a function of the photon energy. This periodicity matches very well that expected for $k_z$. This was recognized before[20,15] but the inner structure of A and B bands was not resolved. In $BaFe_2As_2$, there are 2 FeAs slabs per unit cell, so that the distance between two slabs is $c'=c/2=6.5$Å and the periodicity expected for the perpendicular dispersion is $2\pi/c'$. At the Γ point, the value of $k_z$ can be estimated by :

$$k_z = 0.512\sqrt{h\nu - W + V_0}$$

where hν is the photon energy, W the work function of the material (W~4.4eV) and $V_0$ an inner potential[22]. It is usually of the order of 10eV, but is in practice adjusted to get a reasonable agreement with the data. Using $V_0$=14eV, we obtain $k_Z$=1 for the maxima of $k_F$ at 34eV ($7\pi/c'$), 62eV ($9\pi/c'$) and 98eV ($11\pi/c'$) and $k_Z$=0 for the minima at 22eV ($8\pi/c'$), 48eV ($8\pi/c'$) and 79eV ($10\pi/c'$). This describes very well the experimental minima and maxima, which supports this interpretation.

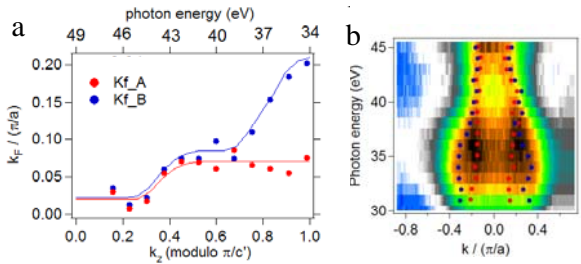

FIG. 4 : Detailed variation of $k_F$ with photon energy for $x=0.045$. (a) $k_F$ for A (red points) and B (blue points) bands extracted with linear fits of the dispersion. Solid lines are guides for the eye. (b) Spectral intensity at a binding energy of –40meV as a function of photon energy. Red and blue points indicate the position of the A and B bands at this binding energy.

The 3D behavior is the same in all the samples we have investigated. Recently, Liu et al. claimed that the 3D structure was characteristic of the orthorhombic (magnetic) phase of $BaFe_2As_2$ and disappeared at high temperature in the tetragonal phase[12]. At the various dopings studied here, the structure changes from orthorhombic to tetragonal, and the properties from magnetic to superconducting to simply metallic, so that if there is a change of dimensionality as a function of temperature, it is not likely associated to structural or magnetic transitions.

A more detailed variation of $k_F$ with $k_Z$ is shown in Fig. 4 for $x=0.045$. Panel (a) shows the positions of the two bands extracted by linear fits as a function of $k_Z$, while panel (b) directly images the shrinking of the cylinders as a function of $k_Z$. We evidence a moderate increase of $k_F$ from $k_Z=0$ to $k_Z=0.6$, followed by a steep increase for band B towards $k_Z=1$. This shows that the band B has a more pronounced 3D character than A. At low $k_Z$, it is difficult to determine whether both bands move to lower $k_F$ or only one. From the qualitative arguments presented in introduction, it is tempting to associate this more 3D band with the third hole pocket, and consequently, the band A with the two folded 2D pockets.

Indeed, calculations for $BaFe_2As_2$ are in good agreement with this assignment[28,29]. In ref. [28], two nearly circular hole pockets are found nearly degenerate around the Γ point, with modest $k_z$ dispersion, while a third hole pocket, also circular, is much more 3D. It is nearly degenerate with the first 2 pockets at $k_z$=0, but increases abruptly near $k_z$=1. This behavior corresponds qualitatively very well to that of band B. Recently, Malaeb et al. reached a similar conclusion in a study at $x$=0 and $x$=0.14 [21]. We note that the radii of the pockets we observe are roughly a factor 2 smaller than in the calculations. We believe this is a true deviation between experiment and theory that we will discuss in section 4.

Very recently, a quite different dependence of the electronic structure towards $k_z$ has been claimed for $CaFe_2As_2$ [12]. In this report, it was assumed that the outer band is nearly independent of $k_Z$, and one of the inner band shrinks to zero for small $k_z$ values. Further work will be needed to clarify whether this structure could depend sensitively on the environment of the FeAs planes or whether matrix-element effects may confuse the interpretation of the data. If the intensity of band B was systematically suppressed near $k_z$=0, we may mistake this extinction for a shift. Experimentally, this is a delicate question. The good agreement with band structure calculations a priori supports our interpretation, although we note there are also calculations predicting less 3D character for the outer hole pocket [30].

## III. SHAPE OF THE ELECTRON POCKETS

The case of electron pockets is more simple, because, as we will show, the effect of $k_z$ dispersion is nearly negligible. On the other hand, their shape is more complicated and has been the subject of some debate. At low dopings, Zabolotnyy et al. measured a "propeller-like structure" for these pockets, with sharp "blades" oriented along the BZ diagonals[14]. They argued that this was due to a large reconstruction of these pockets due to a $(\pi/a,\pi/a)$ interaction with the hole pockets. On the contrary, Yi et al. argued that the complicated structure can be understood from the interactions between 4 bands present in the LDA near the Fermi level [9]. Fig. 5 shows that the shape simplifies when the band fills up and the Fermi level moves away from the bottom of the band. At $x$=0.08, the electron parabola is still very shallow and emerges from a high intensity background only in a 40meV window below $E_F$ (red arrows indicate $E_F$ crossings). At $x$=0.3, it becomes very clear over 150meV. Note that this doping is much higher than the highest doping previously studied ($x$=0.15[18]), which is very helpful to determine the trend in $k_F$ variations (Fig. 7). The Fermi velocity changes from $V_F$=0.38eV.Å at $x$=0.08 to $V_F$=0.76eV.Å at $x$=0.3. This can be simply understood by a shift of $E_F$ near the bottom of the band. Both values correspond to about the same effective mass, $m^*$=3.5$m_e$ ($m_e$ is the mass of the electron). We have quoted similar values of $V_F$ and $k_F$ for the hole pockets in section 2. For holes, m* is comprised between 2 and $4m_e$ depending on the doping, the band A or B, and $k_Z$.

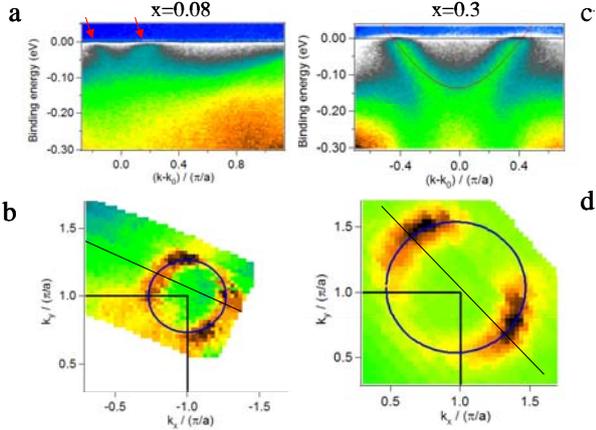

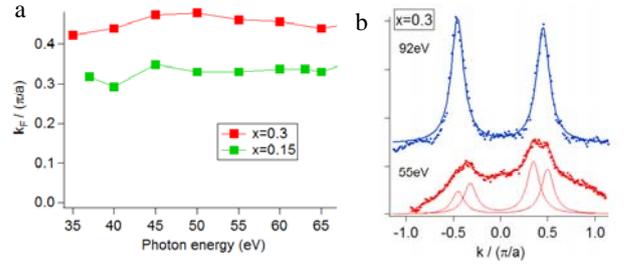

FIG. 5 : Near $E_F$ electronic structure of the electron pockets, for $x=0.08$ (a-b) and $x=0.3$ (c-d). (a) and (c) : Dispersion measured at 30K and 45eV photon energy, in the direction indicated as thin black line on the FS (bottom). (b) and (d) : Fermi Surface measured at 45eV (b) and 92eV (d) for the two different dopings. Circles delimit the FS.

The FS shown in Fig. 5b and 5d are clearly defined, with simple shapes, quite similar at the two dopings (this is identical at $x=0.15$, not shown). They can be reasonably well described by circles, although some segments appear more straight, especially at $x=0.08$. We always observe a higher intensity on the sides of the circle away from the diagonal direction. This is probably due to the fact that these sides have respectively $d_{xz}$ or $d_{yz}$ symmetry and consequently obey different matrix-element effects. The strong dependence of these pockets on the polarization was already noted before[14,17] and is an additional complication to determine the true shape of the electronic pockets.

Fig. 6 shows that we do not observe strong changes of $k_F$ as a function of photon energy. This confirms the 2D character of these bands, similar to that of the hole A bands. We do observe variation of intensity and peak width as a function of photon energy, which may be due to an internal structure of the pockets, with two bands that are not exactly degenerate at all $k_Z$. As an example, Fig. 6b shows spectra near the Fermi level at two different photon energies. At 92eV, the peaks are very sharp, with $0.08\pi/a$ half width at half maximum (HWHM), and centered at $0.43\pi/a$. They broaden at 55eV, but can be well fitted by two peaks of $0.08\pi/a$ HWHM, positioned at $x=0.35\pi/a$ and $0.5\pi/a$. In band calculations, the two electron pockets are degenerate and nearly circular for $k_z=0.5$, but become more oval and less degenerate towards $k_z=0$ and $1$[28,29]. This would explain these variations quite well. Previous studies at $x=0.15$[18] and $x=0.06$[9] attempted fits with two different oval shapes. This would also be possible with these data, but seems to us beyond experimental accuracy. Even though such variations of $k_F$ are not negligible, they do not affect much the average area of the pockets, which is the main quantity we want to determine, so that we neglect them in the following.

FIG. 6 : (a) $k_F$ as a function of photon energy for $x=0.15$ (green) and 0.3 (red). (b) Spectra at -10meV binding energy for $x=0.3$ at 92eV (top) and 55eV (bottom). For 55eV, the decomposition of each peak into two different peaks is indicated by thin lines.

## IV. EVOLUTION WITH DOPING OF $K_F$ AND OF THE NUMBER OF HOLES AND ELECTRONS

Finally, Fig. 7 summarizes the evolution of $k_F$ as a function of doping. For the hole pockets, *maximum values* of $k_F$ are reported (i.e. those for $k_Z=1$). We add results for magnetic phases (x=0 and 0.045), although we do not address the details of the reconstruction of the electronic structure in the magnetic phase in this report. For the hole pockets, we use low temperature data (20K) to be able to resolve the two different bands, but fit the dispersions to linear models, ignoring possible gaps (see Fig. 2). For the electron pockets, we use data taken at temperatures above the magnetic transition to avoid this problem. For low dopings, $k_F$ for the electron pockets is just indicative, as their shape may differ significantly from a circle.

We first note that $k_F$ varies almost linearly with doping. The slopes are quite similar for the hole and electron bands, reflecting the properties of the two types of carriers are not very different. We are not in a situation of heavy holes and light electrons (or the inverse), except, possibly, at some singular points of $k_Z$ and dopings. We add in Fig. 7 two points, measured by [7], corresponding to hole-doped $Ba_{1-x}K_xFe_2As_2$, with $x=0.2$ (0.1 holes/Fe) and $x=0.4$ (0.2 holes/Fe). They extrapolate quite well with $k_F$ determined for the electron-doped side, which strongly suggests to identify the so-called α and β hole pockets[7] with our A and B pockets. In section 2, we argued that the A sheet is probably doubly degenerate and the B sheet singly degenerate, because of its more 3D character. Xu *et al.* also concluded that the inner α sheet must be doubly degenerate and the outer β sheet singly degenerate to satisfy the Luttinger theorem[7]. In our opinion, this correspondence definitely establish these assignments. No $k_Z$ dependence was ever reported for the β band, in contrast with the B band, but it may indeed be more 2D in the K case.

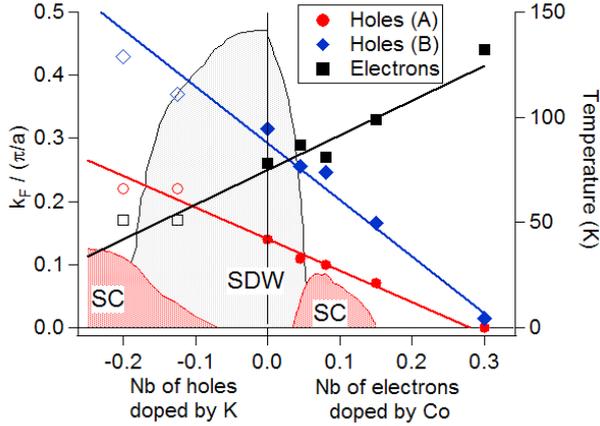
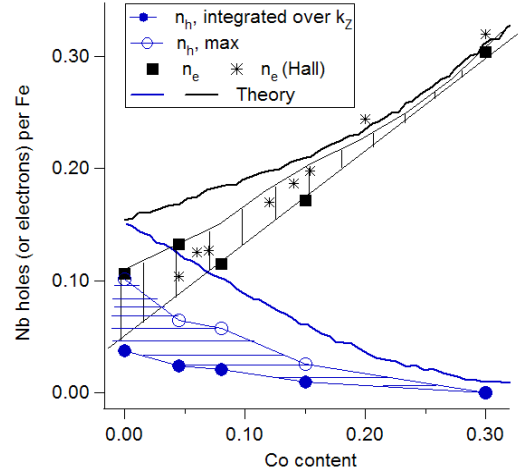

FIG. 7 : Closed symbols : $k_F$ values as a function of the number of doped electrons determined in this study for the electron and A and B hole pockets at $k_Z=1$. Open symbols : $k_F$ values as a function of the number of doped holes, reproduced from ref. [7]. Lines are guide for the eyes. Grey and red hatched areas show SDW and superconducting (SC) regions, respectively (right axis).

FIG. 8 : Number of holes (solid blue circles) and electrons (black squares) extracted from the $k_F$ values of Fig. 7. Open blue circle are maximum values for the number of holes, neglecting $k_Z$ dispersion. Black stars are number of electrons deduced from Hall measurements[27]. Hatched areas indicate likely values for these variations including experimental errors. Thick solid lines are prediction for the evolution of number of holes and electrons reproduced from ref. [30].

To be more quantitative, it is instructive to estimate the number of carriers in the different pockets, as we propose in Fig. 8. In a 2D system, the Luttinger theorem states that the number of carriers contained in one pocket is proportional to its area compared to that of the BZ, even in presence of strong correlations[31]. The BZ is here approximately a square of side $2\pi/a$ with a=3.96Å in the tetragonal phase of $BaFe_2As_2$. This does not change much with Co doping[32]. It contains two holes (or two electrons) and two Fe, hence 1 carrier per Fe. A pocket of area S then contains $S/S_{BZ}$ carriers per Fe in this 2D analysis. In a 3D system, the number of carriers should be integrated over the entire volume, which is possible knowing the evolution of $k_F$ with $k_Z$, as determined in Fig. 4 for the hole pockets.

For the electron pockets, we compute the number of electrons $n_{el}$ by assuming two nearly degenerate, 2D, and circular pockets, yielding $n_{el}=\pi k_F^2/2$. We caution that at low dopings, deviation from a circular shape may be important. The simplest case is that of high Co doping, when the hole pocket is completely filled. In this case, the electron pocket should contain all electrons brought by Co. This is in good agreement with the value $n_{el}=0.29$ that we measure at $x=0.3$. This number decreases with decreasing percentage of Co. However, it seems to converge to a *lower* value than the 0.15 electrons/Fe predicted by ref. [30] at $x=0$. Interestingly, a very similar evolution of the number of electrons with Co doping was deduced from Hall effect measurements on crystals from the same batches (stars in Fig.8), taking the Hall number $|n_H|=1/|R_H|e$ at low temperature as a good estimate for the actual value of $n_{el}$ as explained in ref. This reinforces the idea that the pockets are probably smaller than predicted in theory for the undoped compound. Depending on the extrapolation at low dopings, we estimate $n_{el}(0)=0.07\pm0.03$.

For the hole pockets, the general trend of Fig. 7 extrapolates to $k_F^A=0.14\pi/a$ and $k_F^B=0.28\pi/a$ at $x=0$ and $k_Z=1$. This is significantly smaller than in the calculation of ref. [28], where $k_F^A\approx 0.2\pi/a$ and $k_F^B\approx 0.5\pi/a$ at $k_Z=1$. Consequently, our measurement also corresponds to a smaller number of holes at $x=0$ than predicted theoretically. This confirms the tendency obtained for electrons, which is self-consistent, as charge neutrality requires $n_h(0)=n_{el}(0)$. The $k_F$ values at $k_Z=1$ correspond to ~0.015 holes/Fe in one band A and ~0.06 holes/Fe in band B, adding to a total of 0.09 holes/Fe, assuming double degeneracy of A. This should be taken as a maximum estimate for the number of holes, since the holes cylinders in fact shrink at lower $k_Z$ values, especially for B. Integrating these numbers over $k_Z$ with the dependence determined on Fig. 4, reduces these numbers to 65% for band A (~0.01 holes/Fe) and 25% for band B (~0.015 holes/Fe). Note that almost half the holes are contained in band B, meaning both bands are equally important to consider. The total number of holes is very small [$n_h(0)=0.035$ with this estimation] and this might mean that the integration on $k_Z$ underestimate this number (with these small $k_F$ values, small changes will affect significantly the results). In Fig. 8, we report both the maximum and integrated values, for the different dopings. Comparison with the electron case indicates n=0.06±0.02 as a likely value for the number of carriers at $x=0$. A similar shrinking of hole and electron pockets was observed by de Haas-van Alphen experiments in the non-magnetic

LaFePO[33] and appears therefore as a common feature of pnictide that we will discuss in conclusion.

## V. DISCUSSION OF THE NESTING

One major interest of such a study is to estimate the degree of nesting between the different pockets. As the pockets are roughly circular, they will exhibit good nesting at the wave vector ($\pi/a$, $\pi/a$), if they have similar $k_F$. It is immediately clear from Fig. 7 that a good nesting on the electron doped side will only be found between electron pockets and the B hole pocket in the region $x<0.07$. As this is indeed the region where the SDW are observed, one could conclude that this supports the role of nesting in driving this instability. However, it seems quite counter-intuitive that the 3D pocket alone contributes to the nesting. Also, the good nesting would exist only in a limited $k_Z$ range, due to the strong dispersion of this band. The nesting appears therefore much worse than in the initial naïve 2 band model, in the absence of the 3D pocket. It is also interesting to keep in mind that this third hole pocket seem quite different in other iron pnictides, although the SDW transition temperature are quite similar. $K_F$ can be as large as $0.8\pi/a$ for LaFeOP[8] or $0.5\pi/a$ for NdFeAsO$_{1-x}$F$_x$[10], so that it is unlikely that the SDW critically depends on how it is nested with the electron pocket. Note that this very large pocket has sometimes been assigned to surface effects[33], but that the 3D effects we probe are incompatible with surface effects.

Such a simple view of FS nesting is however probably not very relevant to describe such phases. Johannes and Mazin have shown that the reconstruction of the electronic structure in the magnetic phase extend over almost the entire band width and not just in a small energy window below $E_F$ [34]. Such changes cannot be captured by simple nesting arguments. Indeed, there is no obvious gap opening detected by ARPES at the magnetic transition[15,16], as would be expected for a simple SDW, at least not for the A bands (see Fig. 2). One rather observes a *splitting* of the electrons and holes bands[16] and new bands folded with ($\pi/a,\pi/a$) periodicity[16,17,13], testifying for the new AF BZ boundaries (see Fig. 1). This behavior can also be observed at $x=0$ in Fig. 2, where an electron-like band, shown as white dotted line, becomes visible at the $\Gamma$ point. It is much larger than in the non-magnetic phase ($k_F \approx 0.4\pi/a$), due to the splitting, and interacts with both hole bands when they cross, although this does not result in a straightforward gap at the Fermi level. As a result, the magnetic phase is characterized by a complicated pattern of residual metallic pockets. The origin of the splitting has been discussed in terms of exchange splitting[16] or anisotropic gaps in different domains[17] and is not yet fully clarified, in our opinion. A better observation of the location and sizes of the magnetic gaps would also be needed to fully understand the reconstruction of the electronic structure in the magnetic phase.

It is also interesting to compare the relative sizes of the pockets in the superconducting phase, as interband transitions are supposed to play an active role in superconductivity[4]. Fig. 8 immediately suggests that the disappearance of superconductivity on the electron-doped side coincides with the filling of the hole bands. This was indeed suggested by ref. [19]. More precisely, we have shown in Fig. 3 that the pocket is not yet totally filled, but exists only at particular $k_Z$ values. As it seems very convincing, it is interesting to note that the symmetric behavior on the hole-doped size is not as obvious. There is a very clear asymmetry of $T_c$ on the hole and electron sides. $T_c$ is higher with K doping ($T_{c,max}$=38K) than Co doping ($T_{c,max}$=24K) and also extends on a much wider doping range (up to 0.5 holes/Fe with K doping). If superconductivity also disappears when the two carriers cease to coexist, this means that the electron pocket survives up to $x$=0.5, a much higher doping than on the electron-doped side. This may be possible, if they survive as "blades" [14], containing very few electrons, but still supporting superconductivity. A detailed investigation is likely to reveal interesting aspects on essential ingredients for superconductivity in these systems.

A big difference between the hole and electron sides is that the electron pocket is of the same size as the *inner* pocket for hole doping and as the *outer* pocket for electron doping. As the degeneracy and dimensionality of the inner and outer pockets are not the same, this is quite a different situation. On the hole-doped side, the superconducting gap was found similar on the two pockets of similar size (electron and hole A pockets), but smaller on the B pocket[5]. This agrees very well with the idea that superconductivity is stabilized by interband transitions. Following these ideas, one would expect it to be destabilized on the electron doped case, when the pocket sizes are more different, which might indeed be the reason for the lower $T_c$. At x=0.075, Terashima *et al.* reported a superconducting gap stronger on the hole B pocket than on the electron pocket[19]. They assumed the hole A band was already filled. We show here that it is not the case and a complete investigation of the different gaps as a function of $k_Z$ could reveal interesting aspects of the role of 3D effects and FS nesting in the superconducting properties.

## VI. CONCLUSION

Our investigation clarifies the internal structure of the hole and electron pockets in Ba(Fe$_{1-x}$Co$_x$)$_2$As$_2$. This study is complicated by the overlap between the different bands and strong polarization and photon energy dependence. Disentangling these different effects is however a prerequisite for an insightful study of the evolution of the electronic structure in the magnetic and superconducting states. The way each band participates in these transitions is indeed a key to understand the nature of the ground states. Our study identifies three different hole pockets, the outer one exhibiting strong photon energy dependence. We attribute this to a stronger 3D character of this band, in good agreement with band calculations. We caution that further study of matrix-element effects are required to definitely establish this fact. At large electron doping, we observe 2D electron pockets, with rather circular shapes. The continuity of these structures with the hole-doped side (in $Ba_{1-x}K_xFe_2As_2$) suggests a unified structure, where the third hole pocket always play an important role and the electron pocket survives deep in the hole-doped side because of an increasingly anisotropic shape.

Fig. 7 and Fig. 8 summarize the results of this study. The nesting between hole and electron pockets is not as strong as usually expected, even at $x=0$, and especially when 3D effects are taken into account. On the other hand, the presence of hole and electron pockets seem crucial to stabilize superconductivity. It disappears when the hole pockets are nearly filled and the different $T_C$ on the hole-doped and electron-doped sides may be related to the quite different relative sizes of the pockets. Our study suggests directions to investigate in order to better understand the role of the coupling between hole and electron pockets in superconductivity. It would for example be interesting to compare the size of the third hole pocket, which effectively controls the mismatch between the 2D hole and electron pockets, with the maximum $T_c$ value in the 1111 family, which strongly varies from $T_c$=26K with La to $T_c$=52K with Sm.

The total number of carriers is found in good agreement with the overall bulk stoichiometry. This gives confidence that the measured electronic structure is that of the bulk. On the other hand, we find a number of carriers of each species at $x=0$ that is smaller by about a factor 2 than predicted theoretically[30]. This tendency was already suggested to interpret transport data in $Ba(Fe_{1-x}Co_x)_2As_2$[27] or de Haas-van Alphen oscillations in the non-magnetic LaFeOP[33], so that it is probably an intrinsic feature of these systems. It implies *shifts* of the electronic bands compared to the LDA, not a simple renormalization. This is a consequence of the multi-orbital character of these compounds, which is one of their originality. It will be important to take into account for a correct interpretation of many experiments. The presence of a third hole pocket in most of the phase diagram, promoted by our analysis, also requires similar shifts to be consistent with some of these calculations. It may play an important role in the electronic properties, but is often neglected in minimal modelling of these compounds. The origin of these shifts involves fine tuning of structural parameters and inter-orbital correlations. Ortenzi *et al.* suggested that the shrinking of the pockets could be attributed to strong interband scattering[35] and therefore reveal important feature of the physics of these materials.

---